\documentclass[journal=jacsat,manuscript=article]{elsarticle}
\usepackage{tgadventor}
\usepackage{float}
\usepackage{graphicx}
\usepackage{amsfonts}
\usepackage{amsmath}
\usepackage{url}
\usepackage{multirow}
\usepackage{wrapfig}
\usepackage[T1]{fontenc}
\usepackage{geometry}
\begin{document}
\begin{frontmatter}
\title{A first-principles study on the early-stage corrosion of a NiWNb alloy in a chloride salt environment}

\author{Tyler D. Dole\v{z}al\textsuperscript{a}\corref{cor1}}
\author{Adib J. Samin\textsuperscript{a}}

\affiliation{organization={Air Force Institute of Technology, Department of Engineering Physics},
            addressline={2950 Hobson Way}, 
            city={Wright-Patterson Air Force Base},
            postcode={45433}, 
            state={OH},
            country={USA}}
\cortext[cor1]{tyler.dolezal.1@us.af.mil}
\begin{abstract}
In this work a representative nickel superalloy, Ni$_{70}$W$_{20}$Nb$_{10}$, was investigated in the presence of chlorine to quantify its early-stage dissolution behavior. Surface structures were generated from a bulk configuration sampled in equilibrium using a multi-cell Monte Carlo method for phase prediction. The predicted solid-phase at 800 $^{\circ}$C was Ni$_{72}$W$_{19}$Nb$_{9}$ in a body-centered tetragonal crystal structure closely resembling the Ni$_4$W structure. Chlorine adsorption onto the energetically favored (110) surface showed preference to niobium which acted as a trapping sink on the top surface of the slab model. Our findings suggested that niobium and tungsten enhanced the corrosion resistance of nickel as their presence created regions that were thermodynamically preferred by the incoming chlorine and less susceptible to chlorine-facilitated dissolution from the alloy. Nickel, niobium, and tungsten resisted chlorine-induced dissolution from the surface model up to a 1/3 monolayer coverage of chlorine indicating that all constituents of this alloy possessed superior resistance to localized surface degradation such as corrosive pitting. Further analysis and comparisons between the corrosion resistance of the three metallic species was performed. This work may provide insights that aid in the development of improved structural materials for molten salt reactors. 
\end{abstract}
\end{frontmatter}

\section{Introduction}
A fundamental understanding of material evolution in high temperature corrosive environments is paramount to the field of sustainable energy \cite{misraFluorideSaltsContainer1987}. Technologies in this field depend on a fluid to manage the heat produced and used within the system. Candidate fluids for these applications are molten halide salts whose thermal properties, such as boiling point, thermal conductivity, and specific heat, are well suited for the operational environment (T $>$ 700 $^\circ$C) \cite{manohars.sohalEngineeringDatabaseLiquid2010,sridharanThermalPropertiesLiClKCl2012}. Of these molten salts, chloride salts are gaining in popularity due to their low cost, low viscosities, and the elimination of hydrofluoric acid formation \cite{williamsAssessmentCandidateMolten2006}. More specifically, KCl-MgCl$_2$ is an attractive chloride salt due to its high volumetric heat capacity at 700 $^{\circ}$C (0.46 cal/g - $^{\circ}$C) \cite{williamsAssessmentCandidateMolten2006} and is a proposed coolant for molten salt reactor (MSR) designs. Under the U.S. Department of Energy (DOE) Advanced Reactor Demonstration Program (ARDP) several MSR systems are being developed and explored. Two such systems include TerraPower's molten chloride fast reactor (2016) and Southern Company's Molten Chloride Reactor Experiment (MCRE) in association with TerraPower (2020). Some of the many beneficial features of MSRs over other reactor designs include low-pressure operation, more efficient power conversion, passive heat rejection, and the elimination of solid fuel and the resulting need to build and dispose of it.
\par
A primary issue with MSRs is the corrosion of the salt-facing structural material. At elevated temperatures, corrosion is driven by thermodynamic dissolution of metal-salt compounds into the molten salt coolant \cite{ruhThermodynamicKineticConsideration2006,edeleanuThermodynamicsCorrosionFused1960, keiserCorrosionResistanceNickelBase1976}. Nickel-based superalloys are candidate alloys due to their good high temperature creep strength and resistance to the corrosive mechanisms encountered in the extreme environments of MSR systems \cite{manlyAIRCRAFTREACTOREXPERIMENT1958,kondoHighPerformanceCorrosion2009,marecekCorrosionTestingNickel2015,mortazaviHightemperatureCorrosionNickelbased2022,aiPossibilitySevereCorrosion2019}. The redox potentials for common alloying elements in decreasing order are as follows: W, Nb, Ni, Cr, Fe, Mg, K \cite{guoCorrosionMoltenFluoride2018,taylorIntegratedComputationalMaterials2018}. From a thermodynamic point of view, alloying nickel with W and Nb may be beneficial since these are the two elements with a higher redox potential in molten chloride salts and are thus expected to be less susceptible to dissolution \cite{ludwigHighTemperatureElectrochemistry2011,zhangRedoxPotentialControl2018}. Delpech et al. recently reviewed a wide range of materials under investigation for structural applications in nuclear systems with molten salts \cite{delpechMoltenFluoridesNuclear2010} and suggested alloying with Nb and the replacement of Mo with W, which showed stronger mechanical performance and corrosion resistance at high temperatures. A number of recent investigations \cite{sousaRelationshipNiobiumContent1995,smithEffectNiobiumCorrosion2004,sutherlinCorrosionNiobiumNiobium2005} also concluded that alloying Ni with Nb would be beneficial for MSR applications due to niobium's good corrosion resistance. Prescott et al. and Ai et al. experimentally investigated tungsten's corrosion resistance to Cl and reported negligible degradation in tungsten rich samples \cite{prescottDegradationMetalsHydrogen1989,aiPossibilitySevereCorrosion2019} indicating alloying Ni with W does lead to improved corrosion resistance. Additionally, Ronne et al. \cite {ronne2020} reported the formation of shell-like structures due to the presence of W which bolstered the corrosion resistance of both the pure Ni and NiCr alloy to Cl-induced defects. Nb and W have been experimentally examined together, as well, where it was concluded that the presence of W and Nb enhanced the corrosion resistance of FeNi$_{25}$Cr$_{15}$W$_2$Nb$_2$V$_1$, especially at elevated temperatures \cite{andrianingtyasRoleTungstenNiobium2018}. 
\par 
Supplemental to the recent experimental investigations on structural materials for molten salt systems, computational efforts have provided insight on the atomistic behavior of these systems. Recently, Startt et al. \cite{startt2021} examined the surface vacancy migration in dilute NiCr alloys using first-principle calculations. They reported that vacancy migration faced a lower activation barrier when undergoing an atomic swap with neighboring Cr and that a vacancy would be trapped once it reached the uppermost surface layer. Further investigation is warranted, but it could be that this phenomenon leads to deeper corrosive pitting in the presence of F or Cl in regions where Cr was plentiful and migrated inwards as vacancies `hopped' towards the outer surface layers. Yin et al. \cite{yinFirstprincipleAtomisticThermodynamic2018} examined the preferential dissolution of CrF from the surface of a dilute NiCr alloy and concluded that Cr was thermodynamically favored to dissolve as CrF\textsubscript{3} and CrF\textsubscript{4}. To provide more computational support on the investigation of alloy corrosion resistance specifically against molten salts, we employ density functional theory (DFT) to examine the surface corrosion performance of a representative system, Ni$_{70}$W$_{20}$Nb$_{10}$, in the presence of Cl. The goal of this work is to provide atomistic insight into the early-stage corrosion behavior and address the suitability of alloying Ni with Nb and W for MSR applications from a corrosion resistance point of view. Because the role of Cr in a Ni alloy has been thoroughly reviewed and explored \cite{ruhThermodynamicKineticConsideration2006,delpechMoltenFluoridesNuclear2010,yinFirstprincipleAtomisticThermodynamic2018,yeHightemperatureCorrosionHastelloy2016,wangInfluenceTemperatureGradient2018} we omitted it from our representative alloy.  
\section{Computational Method}
\subsection{Bulk Structure}
The bulk structure was generated using our implementation \cite{dolezalAdsorptionOxygenHigh2022} of the multi-cell Monte Carlo solid-state phase prediction algorithm \cite{niuMulticellMonteCarlo2019, antillonEfficientDeterminationSolidstate2020}, referred to as (MC)\textsuperscript{2}. The simulation was performed at T = 800 $^{\circ}$C and P = 0 bar with three simulation cells each containing 32 atoms. The initial crystal structure of each simulation cell was face-centered cubic (FCC), body-centered cubic (BCC), and BCC, respectively. One Monte Carlo step began with attempting a local flip. This move is described as randomly selecting one of the three simulation cells, then randomly selecting one atom within the chosen simulation cell and ''flipping`` it from its current species type to one of the other two species types. DFT calculations were executed to calculate simulation cell internal electronic energy and volume changes; the settings are discussed below. The acceptance criterion, based on the Metropolis criterion \cite{metropolisEquationStateCalculations1953}, for the local flip is given in Eq \ref{eq:accept-flip}.
\begin{equation}\label{eq:accept-flip}
P^{flip}_{accept} = min\left\{1,\exp\big(-\beta \Delta H + N\Delta G\big)\right\},
\end{equation}
where $\Delta H$ and $\Delta G$ are defined as,
\begin{gather}\label{eq:delta_g}
    \Delta H = m\sum_{i=1}^{m}(U'_{i} + pV'_{i})f'_{i} - m\sum_{i=1}^{m}(U_{i} + pV_{i})f_{i}\\
     \Delta G = \sum_{i=1}^{m}[f'_{i}\ln(V'_{i}) - f_{i}\ln(V_{i})] + \sum_{i=1}^{m}f'^{i}\sum_{j=1}^{m}X^{' i}_{j}\ln(X^{' i}_{j}) - \sum_{i=1}^{m}f^{i}\sum_{j=1}^{m}X^{i}_{j}\ln(X^{i}_{j}).
\end{gather}
Here, $\beta = 1/k_B T$, where $k_B$ is the Boltzmann constant, N is the sum of all the particles across all simulation cells, $m$ is the total number of simulation cells, $U_i$ is the energy of simulation cell $i$, $V_i$ is the volume of simulation cell $i$, $p$ is the pressure, which was set to 0 Pa, and $f_i$ is the molar fraction of simulation cell $i$. Lastly, letting $n^i_j$ be the number of species $i$ in simulation cell $j$, the atomic concentration is given by $X^i_j = n^i_j/\sum_{k=1}^{3}n^k_j$, which is the atomic concentration of species $i$ in simulation cell $j$. The denominator of $X_j^i$ is the total number of atoms in simulation cell $j$ (32 atoms in our case). The primed coordinates indicate post-flipped values. Finally, and in case a "flip" is accepted, the Lever rule was used to enforce mass conservation and update the phase fractions.
\par 
Spin-polarized DFT calculations were performed using the Projector Augmented Wave (PAW) method as implemented by the Vienna ab initio Software Package (VASP) \cite{kresseEfficientIterativeSchemes1996,kresseUltrasoftPseudopotentialsProjector1999}. The calculations were performed with a plane wave cutoff energy of 450 eV and a 3$\times$3$\times$3 Monkhorst-Pack \cite{monkhorstSpecialPointsBrillouinzone1976} k-point mesh. DFT calculations performed on the simulation cells allowed for changes in the volume and atomic positions through the setting ISIF = 3. The electronic self-consistent calculation was converged to 1$\times$10${}^{-6}$ eV and ionic relaxation steps were performed using the conjugate-gradient method (IBRION = 2) and continued until the total force on each atom dropped below a tolerance of 1$\times$10${}^{-2}$ eV/\AA. The generalized gradient approximation (GGA) was used for the exchange correlation functionals as parameterized by Perdew-Burke and Ernzerof (PBE) \cite{perdewGeneralizedGradientApproximation1996}. The PAW pseudopotentials \cite{kresseUltrasoftPseudopotentialsProjector1999} were used with the valence electron configurations 3d${}^8$4s${}^2$, 4p${}^6$5s${}^1$4d${}^4$, and 6s${}^2$5d${}^4$ for  Ni, Nb, and W, respectively. 
\subsection{Surface Study}
\begin{figure}[H]
    \centering
    \includegraphics[width = \linewidth]{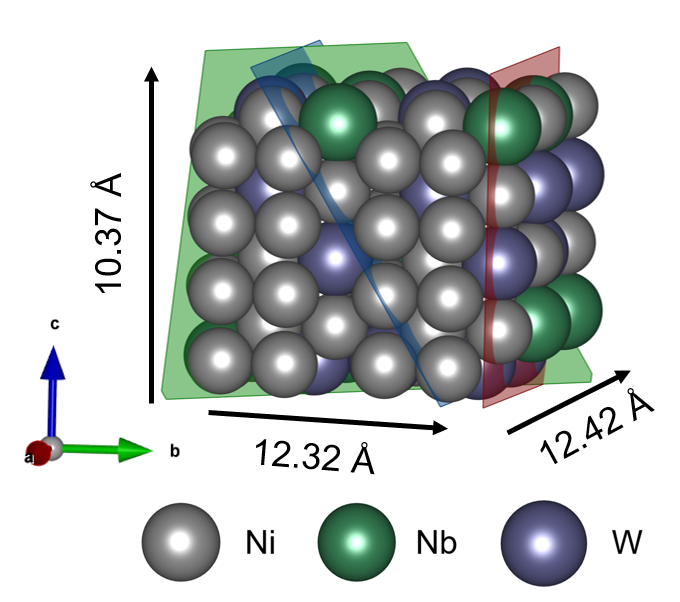}
    \caption{The body-centered tetragonal bulk structure multiplied into a larger supercell for surface cuts. The (010), (011), and (111) surface planes are shown in red, blue, and green, respectively.}
    \label{fig:vesta-faces}
\end{figure}

Once (MC)${}^2$ reached equilibrium, the bulk configuration with the highest molar fraction value was extracted and surface slabs were generated from it. In our case, this corresponded to the 2nd simulation cell, a body-centered tetragonal (BCT) crystal structure with a molar fraction value of 97\% and an atomic concentration of Ni\textsubscript{72}W\textsubscript{19}Nb\textsubscript{9}. Surface slabs were generated along the (100), (110), and (111) directions. The bulk structure and surface cuts are shown in Figure \ref{fig:vesta-faces} (as seen in the Visualization for Electronic and Structural Analysis (VESTA) software \cite{mommaVESTAThreedimensionalVisualization2011}). Surface terminations were chosen such that each slab contained all members of the alloy to maximize insight on how the Cl interacted with different members of the alloy. The surface slab with the lowest surface energy was chosen for additional study. For this work that corresponded to the (110) slab. The surface slabs consisted of four layers, two to represent the surface, and two to represent the bulk, with a total count of 72 atoms, and a vacuum layer of 20 $\textrm{\AA}$ along the $\hat{c}$ direction to prohibit interaction between image slabs. Selective dynamics was used to freeze the bulk layers while the surface layers and Cl-adsorbate(s) were free to move in the $\hat{a}$, $\hat{b}$, and $\hat{c}$ directions.
For relaxation of the clean slab model, the spin-polarized DFT calculations were executed with a plane wave cut-off energy of 450 eV and a 4$\times$4$\times$1 Gamma-point-centered Monkhorst-Pack k-point grid \cite{monkhorstSpecialPointsBrillouinzone1976}. The electronic energy was converged with respect to the k-point grid and energy cut-off to within 1 meV. Through propagation of error in the electronic energies, surface energies were reported with confidence up to 0.1 meV/\AA${}^2$. Electronic relaxation was converged to 1$\times$10${}^{-6}$ eV and ionic relaxation steps were continued until a force tolerance criterion of 1$\times$10${}^{-2}$ eV/\AA\ was satisfied. The PAW pseudopotentials had the same valence electron configurations as stated in the previous subsection, with the addition of an electron valence configuration of 3p${}^5$3s${}^2$ for Cl. The surface energy per unit area is given by,
\begin{equation}\label{eq:surf-energy}
    E_{surf} = \frac{1}{2A}\left(E_{slab} - E_{bulk}\right),
\end{equation}
where A is the area of the surface slab, in units \AA${}^2$, $E_{slab}$ is the total energy of the the surface slab, in units of eV, and $E_{bulk}$ is the energy of the bulk structure from which the surface slab was cut, in units of eV.
\par
\begin{figure}[H]
    \centering
    \includegraphics[width = \linewidth]{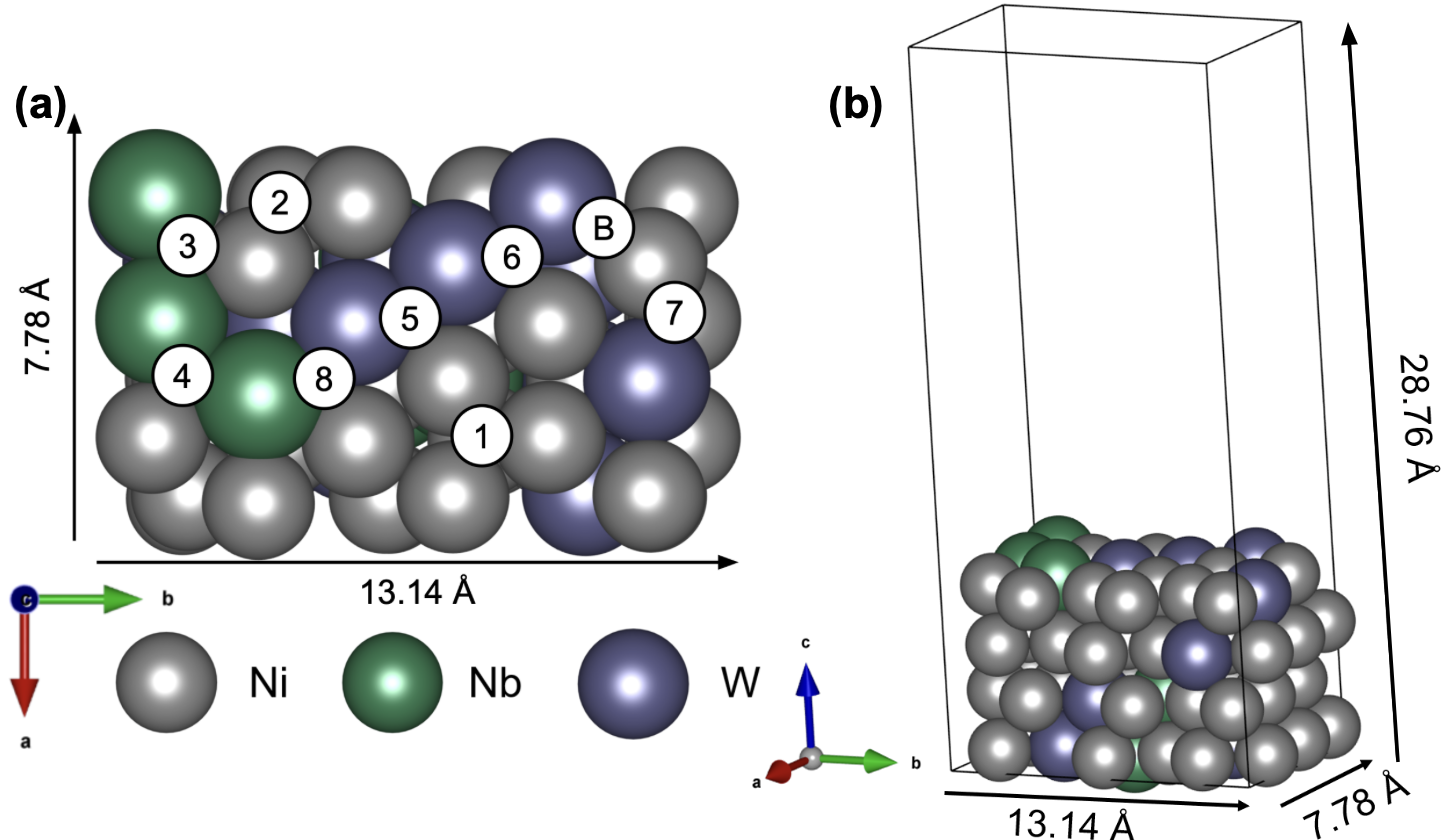}
    \caption{(a) The (110) surface slab with the eight hollow adsorption sites labeled. Because many of the chlorine atoms settled to a bridge adsorption site, one has been marked with with a "B" in the upper right corner of the slab. (b) The surface slab model with the cell bounds displayed.}
    \label{fig:adsorption-and-slab}
\end{figure}

 
For adsorption of a single Cl atom onto the clean (110) surface, the DFT calculations were executed using the same settings as described in the preceding paragraph. Cl was initially adsorbed in the molecular form of Cl$_2$, however it preferred to adsorb in the dissociated form. Similar ab initio investigations of Cl adsorption to metal surfaces observed the same preference \cite{zhaoFirstPrinciplesStudyCl2013, pavlovaFirstPrincipleStudyAdsorption2016,yamashitaFirstPrinciplesStudyChlorine2018}. Therefore, Cl was adsorbed in the dissociated form. Eight adsorption events were carried out in regions of varying Ni, Nb, and W content. Figure \ref{fig:adsorption-and-slab} (a) shows a top-down view of the surface model with the initial placement of the Cl adsorbates marked. Each Cl adsorbate was introduced 3.2 \AA\ above a hollow site. Figure \ref{fig:adsorption-and-slab} (b) is a full view of the slab model with the dimensions labeled and the vacuum region shown.
The adsorption energy is given by \cite{yinFirstprincipleAtomisticThermodynamic2018},
\begin{equation}\label{eq:ads-energy}
    E_{ads} = -\frac{1}{n}\left(E^{Cl(n)}_{slab} - E_{slab} - \frac{n}{2}E_{Cl_2}\right)
\end{equation}
where $E^{Cl(n)}_{slab}$ is the total electronic energy of the slab with adsorbed Cl, $E_{slab}$ is the total electronic energy of the clean slab, $n$ is the number of Cl atoms present, $E_{Cl_2}$ is the energy of the gas phase Cl${}_2$ molecule. The electronic energy of Cl${}_2$ was calculated by relaxing the positions of two Cl atoms in a box of dimensions (12, 12, 12) \AA\ until the force on each atom was below 1x10${}^{-2}$ eV/\AA. The calculated molecular electronic energy was -3.36 eV. To measure the accuracy of this value we calculated the Cl$_2$ binding energy, $BE = E_{Cl_2} - 2E_{Cl}$, and report it to be 2.84 eV which is overestimated when compared to the known value of 2.51 eV \cite{darwentBondDissociationEnergies1970}. This overestimation is attributed to the use of the PBE exchange-correlation functional \cite{droghettiPredictingTextdMagnetism2008,klupfelEffectPerdewZungerSelfinteraction2012}. As defined in equation \ref{eq:ads-energy}, a more positive adsorption energy denotes a more favorable reaction.
\subsection{Corrosion Resistance}
\begin{figure}[H]
    \centering
    \includegraphics[width = \linewidth]{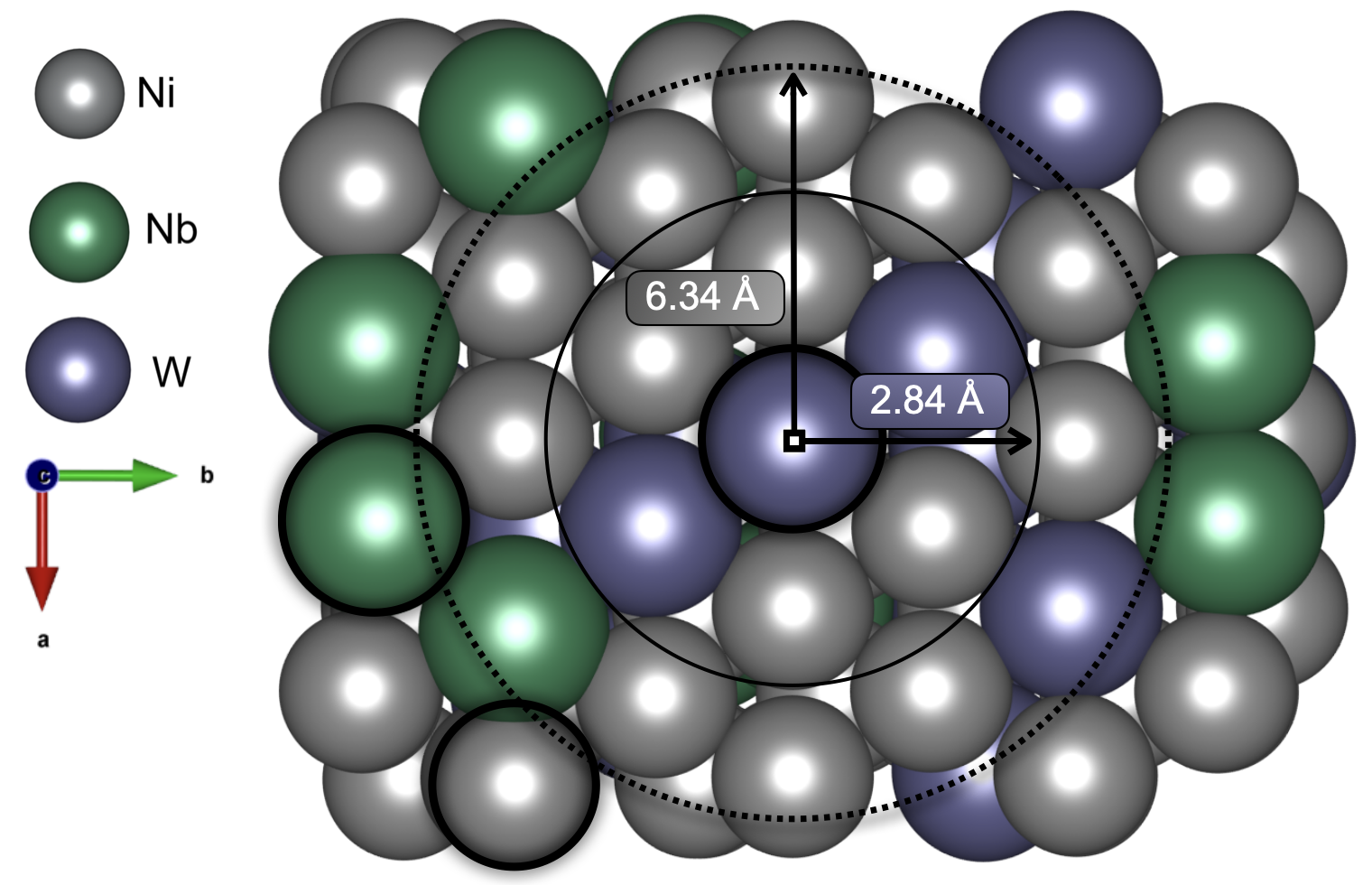}
    \caption{The clean surface slab with the chlorine-attacked niobium, nickel, and tungsten atoms encircled in black. For this example, the solid black and dotted rings indicate the approximate areas that were scanned to introduce attacking atomic chlorine around tungsten. The same procedure was followed for the nickel and niobium attacks.}
    \label{fig:three-atoms}
\end{figure}

To study the mechanisms of attack, Cl was systematically adsorbed around a Ni, Nb, and W atom up to five Cl atoms for a surface coverage of 5/15 monolayer (ML) with freedom to move in all three directions. The DFT calculations were executed with the same settings as described in the preceding section. We considered 1 ML to represent a complete occupation of the 15 available hollow sites on the surface layer. Figure \ref{fig:three-atoms} shows the clean surface slab with the attacked Ni, Nb, and W atoms encircled in black. The solid and dotted ring show the approximate area that was scanned to introduce each atomic Cl adsorbate. For example, before the system state would progress from 0 ML to 1/15 ML, an adsorption energy sweep was performed around the metal atom in question to identify the energetically favored Cl adsorption location nearest to the metal. This remained true from 0 ML to 5/15 ML. Therefore, the energetically favored adsorption sites within the solid ring were occupied first followed by the energetically favored adsorption sites between the solid and dotted ring. The adsorption energies were calculated using equation \ref{eq:ads-energy}. 
\par 
The desorption energy for each attack from 0 ML up to 5/15 ML was calculated using equation \ref{eq:desorption}. To accurately report the favorable desorbate, we considered all possible desorbates with each new Cl adsorbate. For example, with the NiCl$_5$ event, we consider the desorption of, 1) Ni, 2) NiCl, 3) NiCl$_2$, 4) NiCl$_3$, 5) NiCl$_4$, and 6) NiCl$_5$. Additionally, all permutations of molecular Cl and metal-chloride desorbates were attempted to find the combination that was energetically favored to desorb.
\begin{equation}\label{eq:desorption}
    E_{des} = \left(E_{X} + E_{slab-X}\right) - E_{slab+X}
\end{equation}
In equation \ref{eq:desorption}, $E_{X}$ is the electronic energy of the X molecule in vacuum, i.e. Cl$_2$, NbCl, WCl, NiCl, etc., $E_{slab-X}$ is the electronic energy of the desorbed surface slab, and $E_{slab+X}$ is the electronic energy of the slab with the $X$ compound adsorbed. $E_{X}$ energies were calculated by relaxing the $X$ compound in a box of dimensions (15, 15, 15) \AA.
\subsection{Thermodynamic Calculations}
The Gibbs free energy of formation per atom was calculated for the Nb and W attack configurations at 1/15, 2/15, and 3/15 ML using equation \ref{eq:gibbs-energy}. We did not go beyond 3/15 ML because events beyond this would not be studying Cl adsorption to a Nb or W atom. The reason we did not include the Ni attack configurations is because only one Cl atom actually adsorbed to the targeted Ni. The results would not accurately describe the thermodynamics of adsorption \textit{to a nickel atom}. 
\begin{equation}\label{eq:gibbs-energy}
    \Delta G = \left(E_{slab}^{Cl(n)} + F_{vib} - E_{slab} - n\mu_{Cl}(T,P) - TS_{mix}\right)/(N+n)
\end{equation}
In equation \ref{eq:gibbs-energy}, $E^{Cl(n)}_{slab}$ is the total electronic energy of the oxidized slab, $E_{slab}$ is the total electronic energy of the clean slab, $N$ is the number of metal atoms, and $n$ is the number of Cl atoms present. $F_{vib}$ is the Helmholtz vibrational energy which was considered within in the harmonic approximation for the adsorbed Cl atoms only.
The phonon normal mode frequencies were calculated using the finite-difference method as implemented by VASP (IBRION = 5) with $\Gamma$-point sampling only. The term $\mu_{Cl}$ represents the chemical potential of Cl and is expressed as,
\begin{equation}\label{eq:chem-pot}
    \mu_{Cl}(T,P_{Cl}) = \frac{1}{2}E_{Cl_2} + \Delta \mu_{Cl}(T,P_{Cl}). 
\end{equation}
The first term in equation \ref{eq:chem-pot} is the energy of the gas phase Cl$_2$ molecule. The second term ($\Delta \mu_{Cl}$) is a correction term that is treated as a parameter and is expressed as,
\begin{equation}\label{eq:chem-corr}
    \Delta \mu_{Cl}(T,P_{Cl}) = \frac{1}{2}\left(\mu^0_{Cl_2}+k_B T\ln\left(\frac{P_{Cl}}{P^{0}}\right)\right).
\end{equation}
The term $\mu^0_{Cl_2}$ in equation \ref{eq:chem-corr} is the difference of the chemical potential of Cl${}_2$ at T = 0 K and T $>$ 0 K under P = 1 atm and can be calculated at various temperatures (while holding the pressure constant) using the value reported for a given temperature in the National Institute of Standards and Technology (NIST) Joint Army-Navy-Air Force (JANAF) Thermochemical tables for Cl$_2$ \cite{zotero-390,clearyUseJANAFTables2014,saminOxidationThermodynamicsNbTi2021}. The entropy of mixing ($S_{mix}$) in equation \ref{eq:gibbs-energy} is defined in equation \ref{eq:entropy} and is a constant value in this case because the atomic concentrations of each species ($x_i$) are unchanged.
\begin{equation}\label{eq:entropy}
    S_{mix} = -k_B\sum_{i=1}^{3} x_i\ln(x_i)
\end{equation}
For this study, P$^0$ was set to 1 bar and the system was analyzed in the temperature range of 430$^{\circ}$C to 930$^{\circ}$C and the pressure range of 10$^{-40}$ bar to 10$^{-5}$ bar. For each set of (T, P) the coverage with the lowest Gibbs free energy was identified and a stability plot was generated.
\section{Results and Discussion}
\subsection{Bulk Structure and Surface Structures}
(MC)${}^2$ predicted a BCT structure with lattice constants a, b, c = 6.21, 6.16, 3.46 \AA\ and lattice angles $\alpha\textrm{, }\beta\textrm{, } \gamma$ = 89.8, 90.3, 93.3 degrees. The molar fraction of this phase was 97 $\pm$ 0.289\% and the atomic composition was Ni$_{72}$W$_{19}$Nb$_{9}$ (i.e., Ni$_{50}$W$_{40}$Nb$_{10}$ by weight percent) with an uncertainty of 3.125 at.\% at T = 800$^{\circ}$C and P = 0 bar. To examine the skewing of the crystal structure, the (MC)${}^2$ simulation cell was transferred into a perfect BCC structure and relaxed. From this, it was determined that the skewing of the crystal structure was an artifact of the electronic relaxation and varying sizes of the atomic species. To correct for this, the lattice angles were set to 90$^{\circ}$ before surface cuts were generated. A comparison to the Ni-Nb and Ni-W phase diagrams at 800$^{\circ}$C near 20 atom \% Nb and W found that the Ni-Nb alloy exhibits largely an orthorhombic phase, Ni$_3$Nb \cite{tomihisaPhaseRelationMicrostructure2002,hagiharaDeformationTwinsNi3Nb2001}, and the Ni-W alloy exhibits a complete BCT phase, Ni$_4$W \cite{gabrielExperimentalCalculatedPhase1985}. While a ternary phase diagram was not found, the (MC)$^2$ results match well with the Ni$_4$W BCT structure, whose lattice constants are a = b = 5.73 \AA\ and c = 3.55 \AA, when we account for an increase in the lattice size as a result of the varying sizes of Ni, Nb, and W as they occupy different lattice sites.
\par
The BCT structure had a Bulk Modulus (B) of (183, 187) GPa, a Shear Modulus (G) of (96, 273) GPa, and a Poisson's Ratio ($\mu$) of 0.125 which are in good agreement with the values for pure Ni. The Cauchy Pressure (CP) was calculated using the relation $CP = C_{12} - C_{44}$ and found to be 60 GPa. Using the the elastic constant values reported by Chen et al. we calculated a CP value of 48 GPa for NiNb$_6$ and 50 GPa for NiW$_6$ (by at.\%) \cite{chenElasticPropertiesMultiComponent2010}. It has been suggested that the CP of a system could describe the bonding behavior, where a positive value suggests metallic bonding, a negative value suggests covalent bonding, and a near-zero value indicates a system governed by a central force potential \cite{johnsonAnalyticNearestneighborModel1988}. Based on that, we report good metallic bonding behavior in the alloy system and see good agreement between our ternary alloy results and the binary alloy results. Lastly, the surface energy values for the (100), (110), and (111) surfaces were 187.6, 148.6, and 154.7 meV/\AA$^2$, respectively. The adsorption study was carried out on the energetically favored (110) surface slab.
\subsection{Chlorine Adsorption}
\begin{figure}[H]
    \centering
    \includegraphics[width = \linewidth]{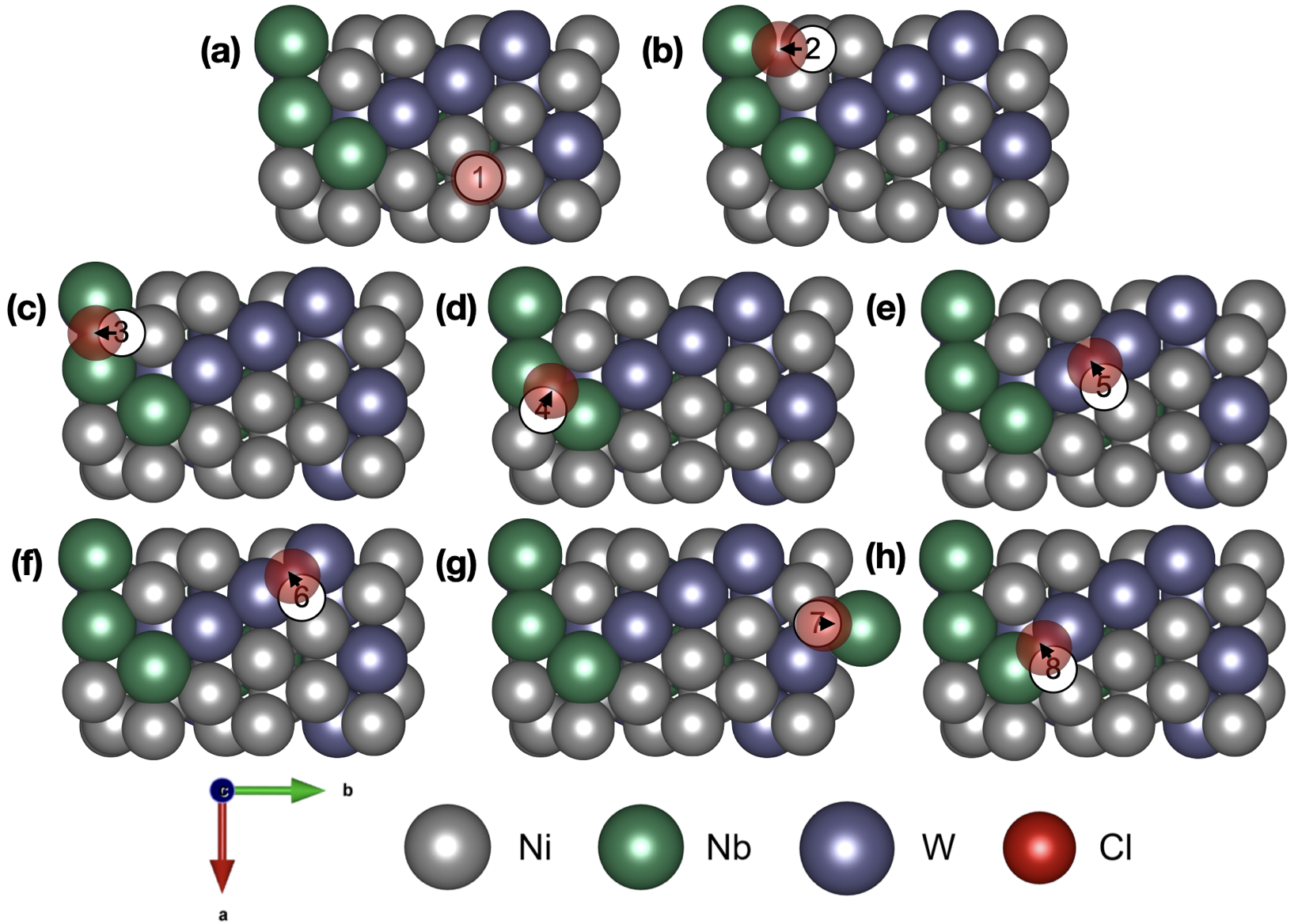}
    \caption{The final chlorine positions for the eight adsorption events with the initial positions labeled and a vector guiding the reader towards the final position.}
    \label{fig:ads-sites}
\end{figure}

\begin{figure}[H]
    \centering
    \includegraphics[width=\linewidth]{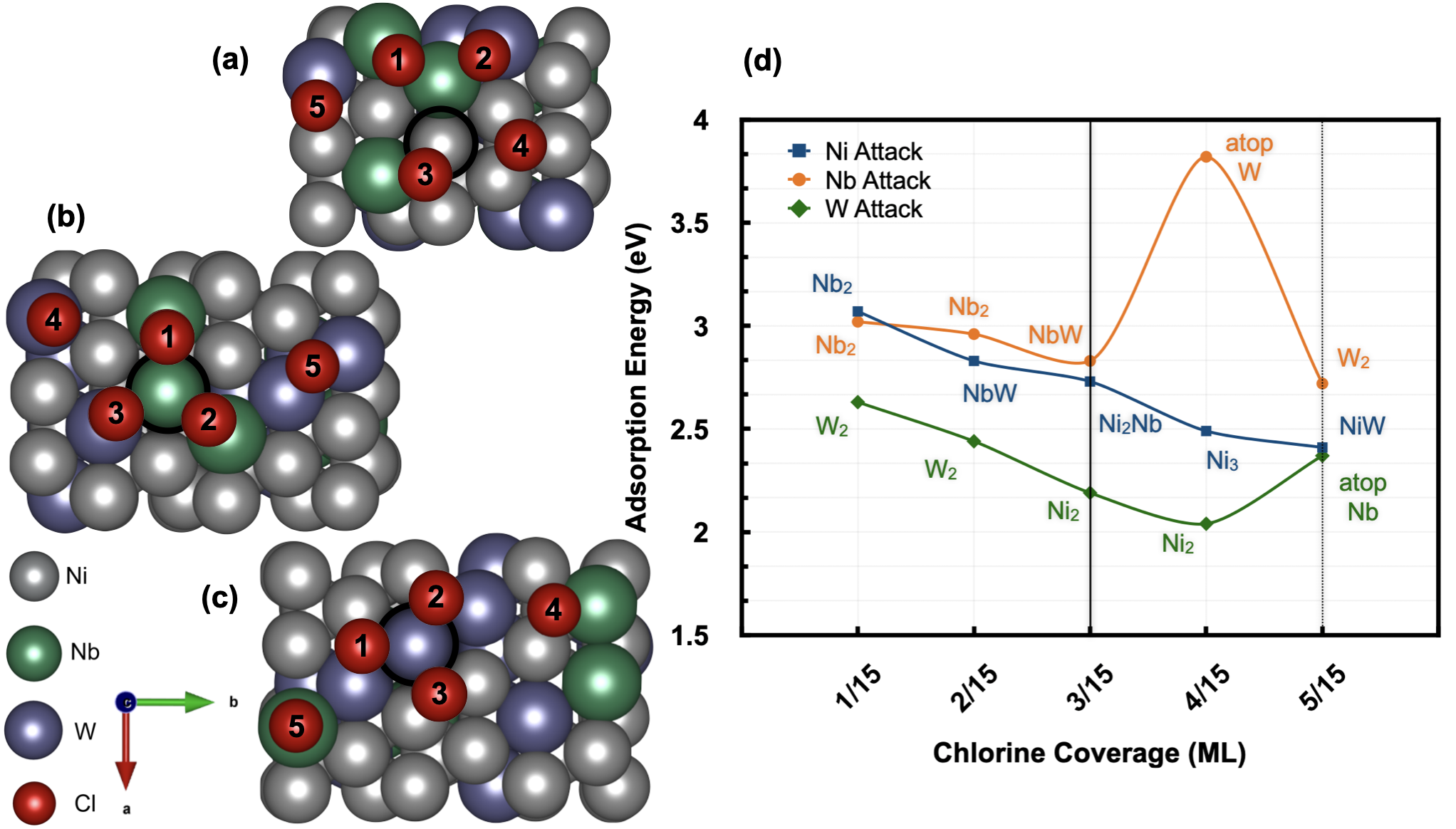}
    \caption{The final surface structure of the (a) nickel, (b) niobium, and (c) tungsten chlorine attacks with the attacked metal atom circled in black. The Cl atoms are labeled according to the order at which they were introduced. (d) The adsorption energies for the events depicted in (a) blue, (b) orange, and (c) green with the nearest metallic neighbors labeled. The solid and dotted black vertical lines indicate the boundaries of the search rings shown in Figure \ref{fig:three-atoms}.}
    \label{fig:adsorption}
\end{figure}

Figure \ref{fig:ads-sites} displays the final state of the eight adsorption events relative to the initial position. The adsorption energies for the eight events are listed in Table \ref{tab:ads-energy} along with the metallic neighbors present at the adsorption site. Two metals listed indicates the Cl relaxed to a bridge site while three metals listed indicates Cl relaxed to a hollow site.
Referencing Table \ref{tab:ads-energy}, Cl preferred the region rich in Nb where the highest adsorption energies were reported. As shown in Figure \ref{fig:ads-sites}, Cl relaxed to sites with Nb present in event (b), (c), (d), (g), and (h). When Nb$_2$ or W$_2$ was present at the adsorption site Cl settled at the bridge site. The lowest adsorption energy was observed at a pure Ni hollow site and the average adsorption energy on this surface was 2.65 eV.
\begin{table}[H]
    \centering
    \caption{The adsorption energy values at 1/15 ML for the events displayed in Figure \ref{fig:ads-sites} and listed in the same order. Three metals listed indicates a hollow site while two atoms indicates a bridge site. A more positive value indicates a more favorable reaction.}
    \begin{tabular}{c|cccccccc}\hline
       \textbf{Ads. Energy (eV)} & 1.94  &  2.46 & 3.02 & 3.07 & 2.63 & 2.51 & 2.67 & 2.97 \\
       \textbf{Nearest Alloy Neighbors} & Ni$_3$ & Ni\textsubscript{2}Nb & Nb\textsubscript{2} & Nb\textsubscript{2} & W\textsubscript{2} & W\textsubscript{2} & NiNbW & NbW \\\hline
    \end{tabular}
    \label{tab:ads-energy}
\end{table}
The progression of the Ni, Nb, and W attacks are shown in Figure \ref{fig:adsorption} (a), (b), and (c), respectively. For the Ni attack (Figure \ref{fig:adsorption}a) the first and second Cl relaxed ``northwards'' (-$\hat{a}$ direction) and settled into a Nb$_2$ and NbW bridge site, respectively. The third Cl adsorbed to a Ni$_2$Nb hollow site. The fourth and fifth Cl adsorbed to a Ni$_3$ hollow site and NiW bridge site, respectively. Considering the 1/15 ML and 2/15 ML adsorption events of the Ni attack, this marks the fourth and fifth time we have reported relaxation of Cl to Nb sites. It is clear that Nb acted as a trapping sink on this surface. For the Nb attack (Figure \ref{fig:adsorption}b) the first and second Cl adsorbed to Nb$_2$ bridge sites and the third Cl adsorbed to a NbW bridge site. The fourth and fifth Cl adsorbed atop of W atom and to a W$_2$ bridge site, respectively. For the W attack (Figure \ref{fig:adsorption}c) the first and second Cl adsorbed to a W$_2$ bridge site and the third Cl adsorbed to a Ni$_2$ bridge site. The fourth Cl adsorbed to a Ni$_2$ bridge site and the fifth Cl relaxed atop a Nb in response to repulsion from the four Cl atoms previously adsorbed to the surface. 
\par 
The adsorption energies for the events depicted in Figure \ref{fig:adsorption} (a), (b), and (c) are shown in Figure \ref{fig:adsorption} (d), where a more positive value indicates a more thermodynamically favorable process. The solid and dotted black line correspond to the boundaries of the solid and dotted search rings depicted in Figure \ref{fig:three-atoms}. Several observations can be gleaned from examining the adsorption energies. First, adsorption to Nb was favored over W or Ni at 1/15 ML, supporting our earlier result from the eight adsorption events, but also at 2/15 and 3/15 ML. Please note, the first two adsorption events of the Ni attack (blue square markers in Figure \ref{fig:adsorption}d) were to a Nb$_2$ site and NbW site, respectively. Second, that Cl aggregation in the Nb-rich region was favored over Cl adsorption to ``clean'' W or Ni up to 3/15 ML. If we consider the Cl adsorption events within the solid black search circle, all three metal attacks exhibited a decrease in adsorption energy with Cl adsorption to W showing the quickest descent ($\Delta E_{ads}$ of about 1 eV). In contrast, Cl adsorption to Nb slowly decreased ($\Delta E_{ads}$ of about 0.25 eV). This indicates that aggregation of Cl around W or Ni is less thermodynamically favorable than Cl aggregation around Nb which is in good agreement with our observation that Nb acted as a trapping sink on this surface. The large increase in the adsorption energy of the 4/15 ML Nb attack event (Figure \ref{fig:adsorption}d 4/14 ML orange data point) was the result of Cl relaxing atop a nearby W atom in response to Cl-Cl repulsion from the already adsorbed first, second, and third Cl atoms.
The average perpendicular movement (along the $\hat{c}$ direction) of each metal species which were bound to one or more Cl atoms was calculated across all three attack events. The average perpendicular movement of Ni, Nb, and W was 0.03262 \AA, 0.2595 \AA, and 0.1415 \AA, respectively. Five out of the eleven instances of Ni-Cl bonds forming resulted in Ni moving inwards (-$\hat{c}$ direction) while Nb and W only moved outwards (+$\hat{c}$ direction). A likely scenario was that the stronger response between Nb, W, and Cl promoted outward movement of the Nb and W atoms which caused the neighboring Ni atoms to relax inwards. Overall, Nb showed the strongest response to the adsorbed Cl followed by W and Ni. The average distance between the first and second surface layer at 5/15 ML showed an increase, relative to the initial separation at 0 ML, by 0.01078 \AA, 0.02531 \AA, and 0.01780 \AA\ for the Ni, Nb, and W attacks, respectively.
\par
\begin{figure}[H]
    \centering
    \includegraphics[width=\linewidth]{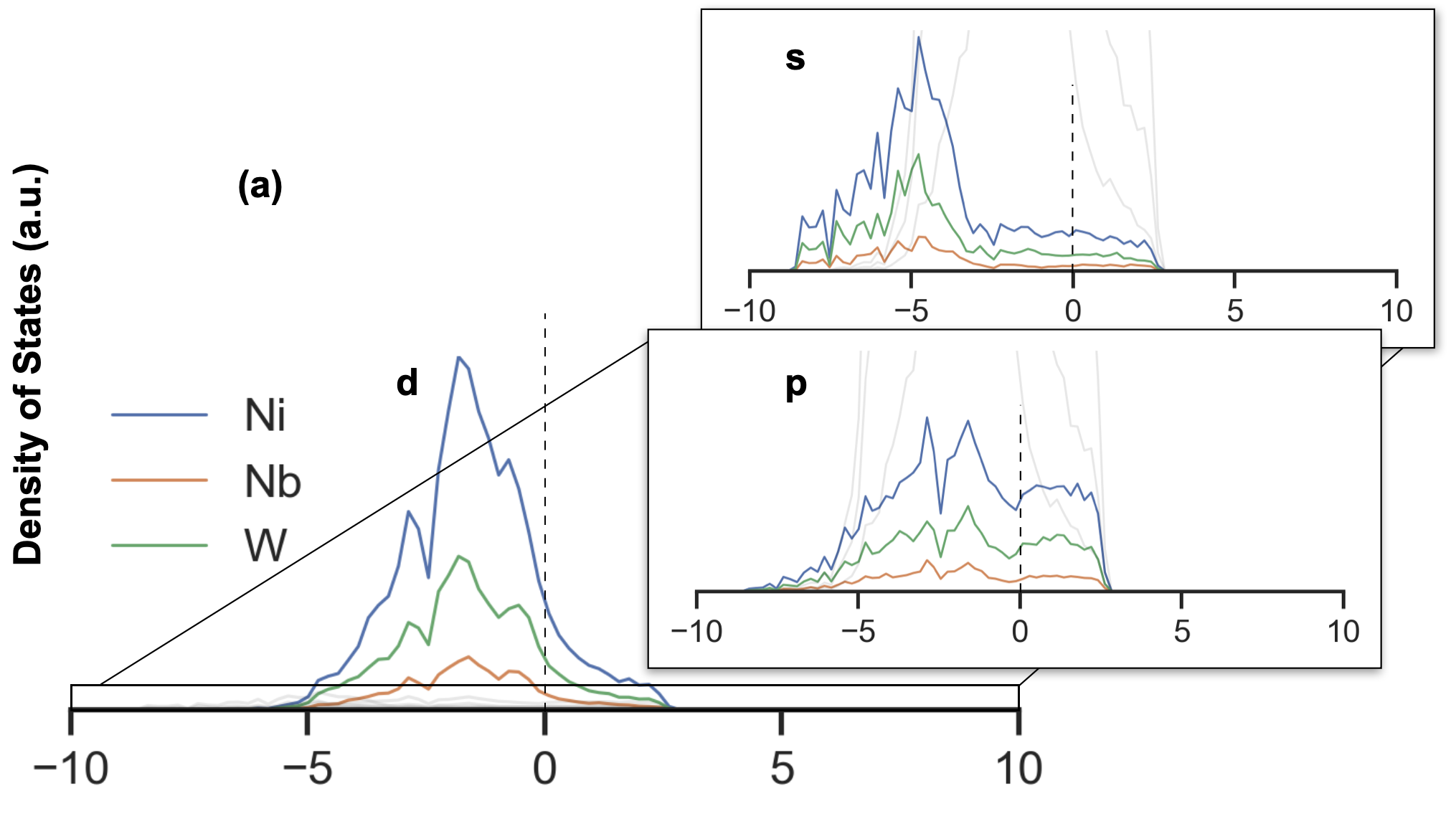}\\
    \includegraphics[width=\linewidth]{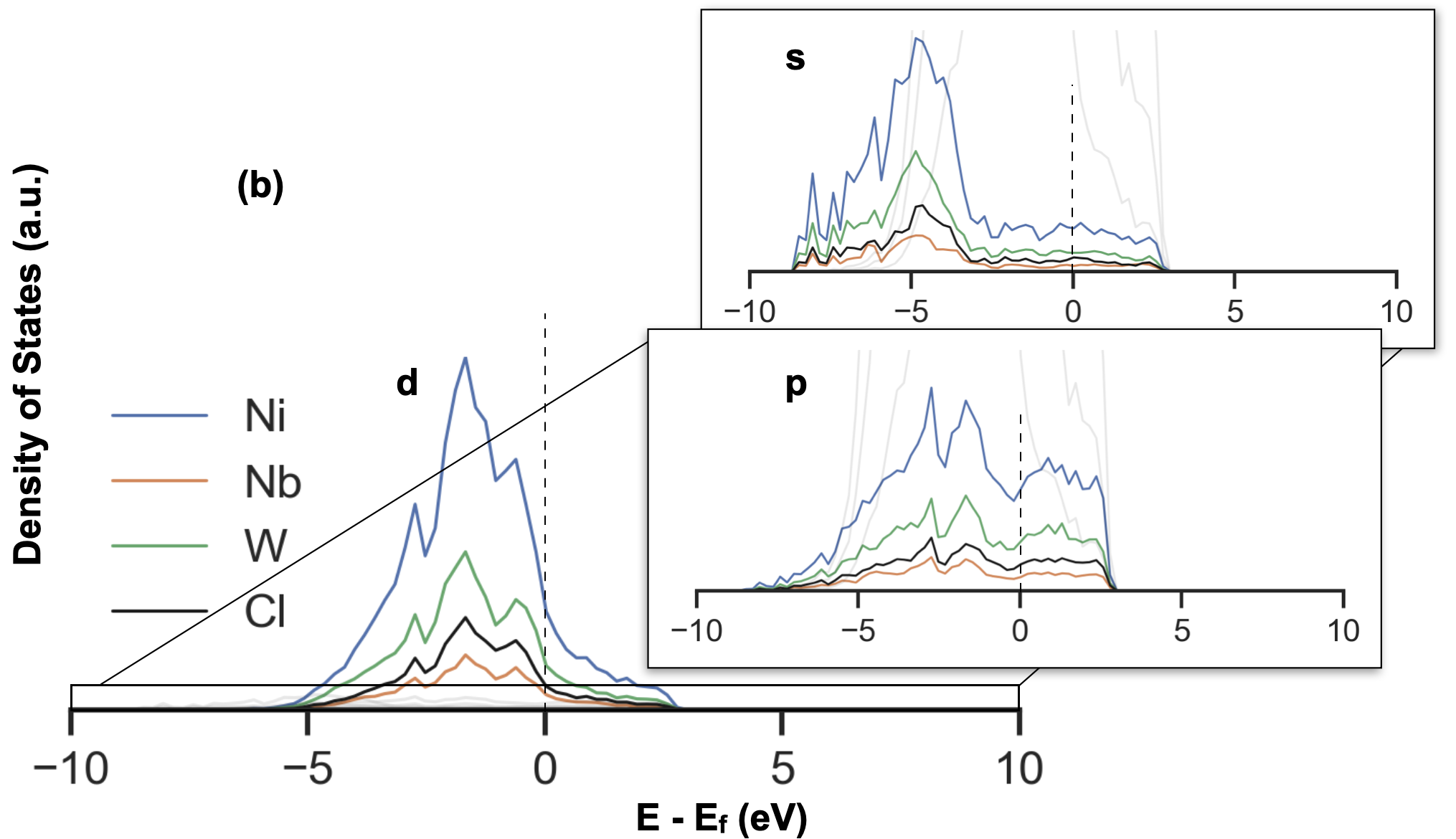}
    \caption{The projected density of states for the nickel (Ni), niobium (Nb), tungsten (W) alloy separated by orbital and color-coded by atomic species for (a) the clean surface slab and (b) the surface slab at 5/15 ML from the Nb attack series with the Cl curve included. The units for the density of states were arbitrary in this case and the scaling for the d-orbital plots was set to include the entirety of the Ni peak. The scaling for the s- and p-orbital graphs was set to include the entire Nb s-orbital peak. To aid the reader with scaling, the s- and p-orbitals were included as gray curves in the d-orbital graph. Similarly, the d-orbital curve was included in gray in the s- and p-orbital graphs.}
    \label{fig:PDOS}
\end{figure}

Figure \ref{fig:PDOS} shows the projected density of states (PDOS) for (a) the clean surface slab and (b) the surface slab at 5/15 ML Cl coverage using the final configuration from the Nb attack sequence. The PDOS curves are separated by the orbitals and color-coded for each metallic species with the addition of the Cl PDOS curve for Figure \ref{fig:PDOS} (b). The s- and p-orbitals are included in the d-orbital graph as gray colored curves to give the reader an idea of scale between the magnitude of states available in the s-, p-, and d-orbitals. Similarly, in the close-up graphs of the s- and p-orbitals the d-orbital has been included as a gray colored curve. From Figure \ref{fig:PDOS}, we see that the majority of the available states resided in the d-orbital near the Fermi level and hybridization amongst the d-, s-, and p-orbitals did occur. A PDOS analysis tells more on the formation of strong metallic bonds between the alloy constituents as the large number of available d-orbital states would promote metallic bonding along with hybridized s- and p- orbitals to increase the cohesion among the metal atoms. This analysis is supported by our previous calculation of the alloy bulk structure's CP which indicated strong metallic bonding. Comparing Figure \ref{fig:PDOS} (a) and (b) against one another shows that the chlorinated surface did remain hybridized to include the newly adsorbed Cl and its electron orbitals. Examination of the PDOS curves adds complementary electronic insight alongside the thermodynamic results based on the adsorption energies of Cl. The strong hybridization of the orbitals explains both the high adsorption energies to the alloy surface and the strong resistance of the alloy against desorption (discussed in Section 3.3) as the adsorbed Cl was tightly bound to the alloy surface and the metal atoms were tightly bound to one another. There was no significant change to the Fermi energy or the overall magnitude and behavior of the PDOS curves between the clean slab and the chlorinated slab suggesting that the electronic properties of the surface slab remained mostly unperturbed up to a Cl coverage of 5/15 ML. This would likely change as more surface sites became occupied by Cl adsorbates and the metallic surface continued to donate electrons causing a growing positive shift in the Fermi energy leading to the eventual desorption of a metal-chloride molecule.
\subsection{Chlorine Desorption}
\begin{figure}[H]
    \centering
    \includegraphics[width=\linewidth]{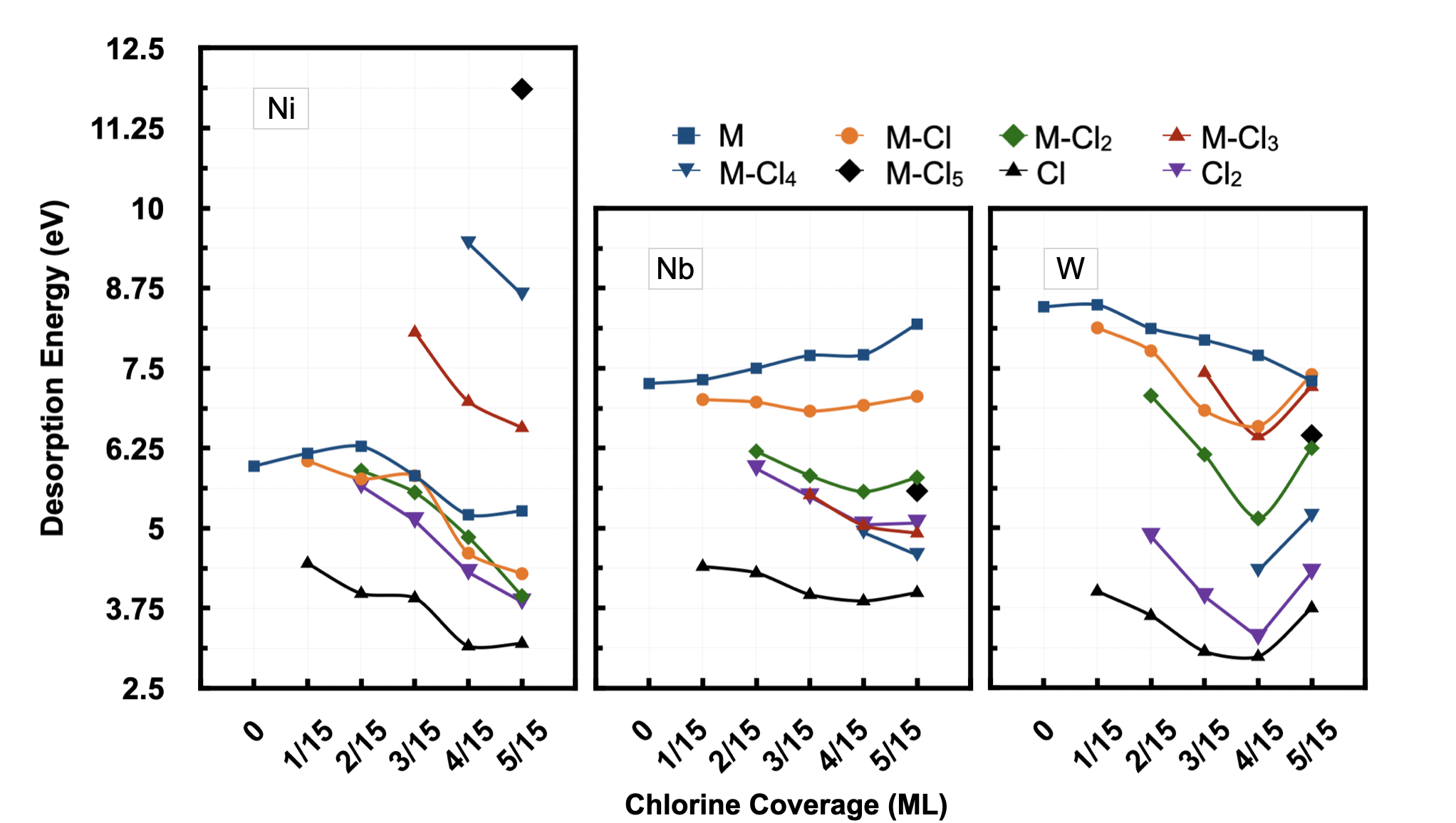}
    \caption{Desorption energies for metal (M) atoms and metal-chloride (MCl$_X$) molecules, where M = Ni, Nb, W and X = 1, 2, 3, 4, 5. Atomic chlorine, Cl, and molecular chlorine, Cl$_2$, desorption energies are listed, as well.}
    \label{fig:desorption}
\end{figure}

To further study the corrosivity of the surface in the presence of chloride salts, desorption energies for the different metal (M), metal-chloride (M-Cl), and Cl atoms and molecules were calculated and are reported in Figure \ref{fig:desorption}, where a lower value indicates a more energetically favored process. First, regardless of which metal was attacked, the favored desorbate was atomic Cl. This fact is clear in Figure \ref{fig:desorption} (a), (b), and (c) where Cl (black line with triangles) is the lowest curve on the plot. The values reported in Figure \ref{fig:desorption} correspond to the atom or molecule which generated the lowest desorption energy. As Cl content increased so did the number of permutations that we needed to scan for potential desorbates. The Cl and Cl$_2$ desorption energies decreased with increased Cl coverage indicating that additional Cl adsorbed nearby encouraged atomic and molecular Cl desorption. Cl and Cl$_2$ were predicted to desorb from Ni or W more readily than from Nb as the desorption energies for Cl and Cl$_2$ were lowest for the Ni and W attacks. Additionally, the 5/15 ML W-attack event (Figure \ref{fig:desorption}c) was Cl adsorbing atop a Nb atom which caused an increase in the desorption energies for any Cl or WCl species ($\Delta E_{des}$ between 0.5 and 1 eV). This suggests that the probability of desorption for a M-Cl molecule is reduced once local NbCl bonds are formed. It may also indicate that forming NbCl bonds protects nearby metal atoms from Cl-assisted dissolution. These observations further support the thermodynamic likelihood that Cl will aggregation in Nb-rich regions. 
\par 
In this study, Ni, Nb, and W resisted Cl-facilitated dissolution from the surface slab up to 5/15 ML, as all metal-chloride curves in Figure \ref{fig:desorption} remained above the Cl curve. This is to say, from a thermodynamic perspective, it is more favorable for Cl to desorb from the metal surface without a metallic partner than with one. It should be noted that in this work, once an M-Cl bond was formed it did not break (up to 4 Cl atoms surrounding the metal partner).
\par 
The magnitude and behavior of the different M-Cl desorption energy curves provides interesting insights. For Ni, its most susceptible dissolution pathway was in the form of NiCl$_2$ followed by NiCl. NiCl formed with three or more Cl (NiCl$_{\geq 3}$) was predicted to be highly resistant to Cl-facilitated dissolution which sets Ni apart from Nb and W as they were more susceptible to dissolution as local Cl content increased. For Nb, its most susceptible dissolution pathway was in the form of NbCl$_4$ followed by NbCl$_3$. We also found Nb had the strongest outward response once NbCl bonds were formed. For W, its most susceptible dissolution pathway was WCl$_4$ followed by WCl$_2$. Referencing the 5/15 ML event in Figure \ref{fig:desorption} (c), we observed yet another example of Nb-attributed corrosion resistance. The sudden increase in desorption energy was a result of the attacking Cl atom binding to a neighboring Nb atom. Comparing the lowest M-Cl data point at 5/15 ML between all three metals shows that NiCl$_2$ ($E_{des}$ about 4 eV) was the most susceptible to Cl-facilitated dissolution, followed by NbCl$_4$ ($E_{des}$ about 4.5 eV) and WCl$_4$ ($E_{des}$ about 5.25 eV thanks to a nearby Nb-Cl bond). Despite the fact that no metallic dissolution was predicted up to 5/15 ML, it is expected that metal-chloride desorption would become more favorable than atomic or molecular Cl desorption as local Cl coverage continued to increase. 
\par 
A similar first principles study was conducted on the adsorption of fluorine to the Ni(111) surface doped with Cr \cite{yinFirstprincipleAtomisticThermodynamic2018}, which has been experimentally confirmed to undergo intense dissolution in the presence of a fluoride and chloride molten salt at MSR temperatures \cite{ruhThermodynamicKineticConsideration2006,yeHightemperatureCorrosionHastelloy2016,wangInfluenceTemperatureGradient2018}. The authors reported CrF adsorption energies between 3 and 3.5 eV/adsorbate and favorable dissolution of Cr in the form of CrF$_2$ and CrF$_3$ whose desorption energies were about 3 and 2 eV, respectively. Pavlova et al. executed a DFT study on the adsorption and desorption of Cl onto the Cu(111) surface \cite{pavlovaFirstPrincipleStudyAdsorption2016}. The authors reported Cl adsorption was favorable on the Cu(111) surface, with adsorption energies around 2 eV/adsorbate. The favored desorbate was CuCl, with an average desorption energy of 2.68 eV. 
\par 
Nickel's superior corrosion resistance is well-known and has been experimentally measured \cite{marecekCorrosionTestingNickel2015,mortazaviHightemperatureCorrosionNickelbased2022}. Ai et al. experimentally demonstrated the superior corrosion resistance of W in the alloy NiW$_{26}$Cr$_6$ where they report only slight depletion in the region rich with W in the presence of FLiNaK molten salt \cite{aiPossibilitySevereCorrosion2019}. Prescott et al. experimentally examined several different metals exposed to an argon-25\%H2-10\%HCl-5\%CO-1\%CO2 gas at 900 $^o$C and concluded Mo and W showed the strongest resistance against Cl-assisted corrosion effects \cite{prescottDegradationMetalsHydrogen1989}. Smith and Eisinger experimentally explored the corrosion resistance of Ni-Cr-Mo-W with the addition of Nb (up to 4 wt.\%) in the presence of different aqueous solutions at T $>$ 650$^{\circ}$C and reported excellent performance against HCl at most concentrations and temperatures; but very poor resistance to HF which is important to note for MSR designs which utilize F-containing molten salts \cite{smithEffectNiobiumCorrosion2004}. Andrianingtyas et al. doped austenitic stainless steel with Nb and W and reported that samples with Nb and W present displayed the highest resistance to pitting corrosion, which increased at elevated temperatures \cite{andrianingtyasRoleTungstenNiobium2018}. 
\par 
Combining the experimental observations on W and Nb corrosion resistance with the magnitude of theoretical adsorption and desorption energies reported by Yin et al. and Pavlova et al., our findings support the experimental conclusions that Nb and W exhibit strong resistance to Cl. Furthermore, from our reported adsorption energies for Ni compared against Nb and W, we found that Cl has a thermodynamic preference to adsorb to Nb (followed by W) over Ni. Couple this observation with the reported desorption energies, which showed W (followed by Nb) possessed a stronger resistance to Cl-assisted dissolution than Ni, it is suggested that Nb and W enhanced the corrosion resistance of Ni. From a thermodynamic point of view, it appeared that Nb and W protected Ni from Cl adsorption by creating more favorable adsorption regions (especially true of Nb) which are more resistant to metal-chloride dissolution (up to 5/15 ML). These observations on Nb and W protection are consistent with experimental studies mentioned in the previous paragraph as well as with studies on Ni-containing ferritic stainless steels which reported increasing Nb \cite{sousaRelationshipNiobiumContent1995} and W \cite{hauganEffectTungstenPitting2017} content reduced the severity of Cl facilitated corrosive pitting in the alloy samples.
\subsection{Thermodynamic Calculations}
\begin{figure}[H]
    \centering
    \includegraphics[width=\linewidth]{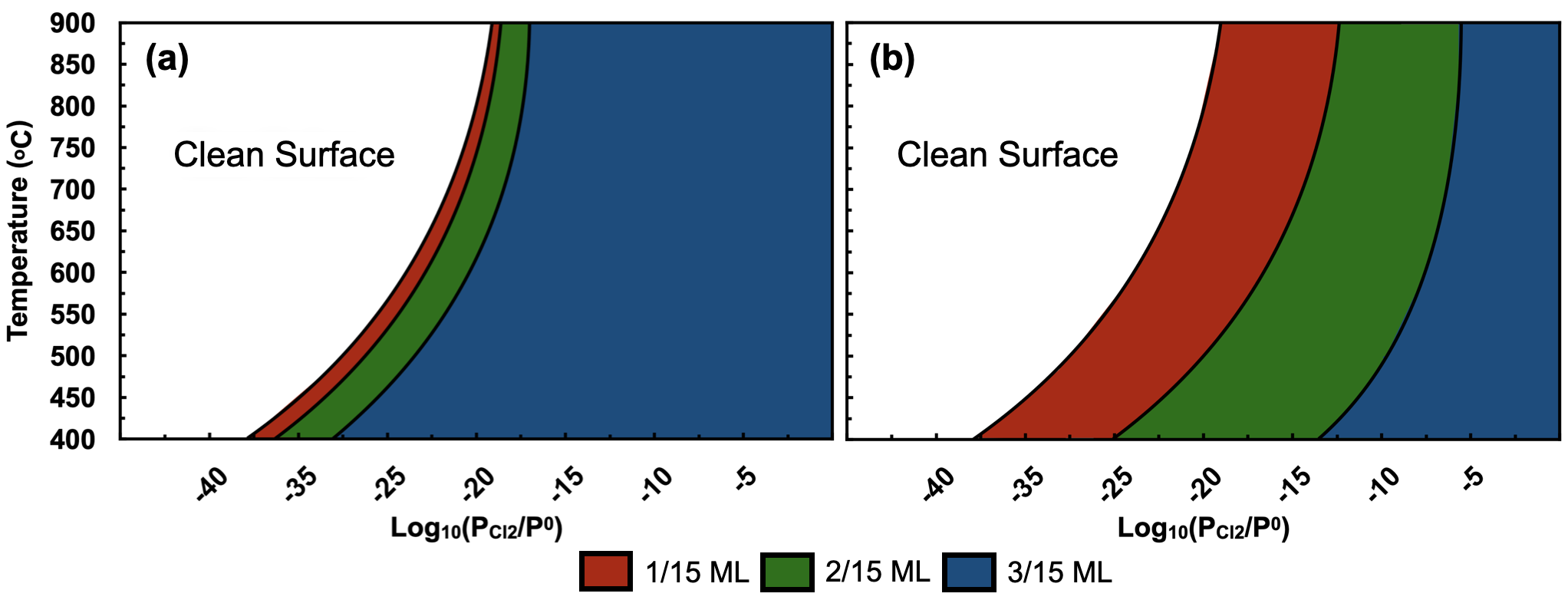}
    \caption{(a) Surface stability plot examining chlorine adsorption directly to a niobium atom. (b) Surface stability plot examining chlorine adsorption directly to a tungsten atom.}
    \label{fig:stability}
\end{figure}
Since Cl adsorption is the first step towards activating the dissolution of metallic atoms, it was important to study the thermodynamic stability of Cl adsorption. The Gibbs Free Energy was calculated for Cl adsorbed directly to a Nb or W atom (up to 3/15 ML) at different (T, P) values. The thermodynamic stability graphs are shown in Figure \ref{fig:stability} where (a) is for Cl adsorbed directly to Nb and (b) is for Cl adsorbed directly to W. Comparing the curves against one another shows that 3/15 ML Cl coverage was predicted to be the most stable structure far more quickly when Cl was adsorbed to Nb over Cl adsorbed to W. Near 800$^{\circ}$C, the WCl plot did not reach 3/15 ML until around $10^{-7}$ bar, while the NbCl plot reached 3/15 ML by $10^{-17}$ bar. Therefore, the presence of Nb increased the thermodynamic stability of the chlorinated surface. This result is in good agreement with our results on Nb from studying the adsorption energies in Figure \ref{fig:adsorption} (d) and could further explain why Cl and M-Cl desorption were thermodynamically less likely to occur as local Nb-Cl bonds were formed.  
\par 
The findings of this work are a first step towards uncovering a deeper understanding on the atomisitic behavior of structural materials in a molten salt environment. The results reported here are in good agreement with the experimental findings, but this study was conducted with limitations. The surface analysis was restricted to a single (110) surface, so results should not be generalized. A more thorough investigation would include additional (110) structures and examine surface models from the \{100\} and \{111\} family. Vibrational corrections were not considered for the surface slab or desorbates during the desorption calculations. Also, anharmonic vibrational corrections to the metal atoms and Cl adsorbates were not applied and could play an important role. At elevated temperatures, desorption of the species considered within this work would likely be more thermodynamically favorable when considering the vibrational contributions. The simulation cell executed in (MC)$^2$ was only a 4$\times$2$\times$2 supercell, restricting the size of the surface structures studied here. A larger surface slab is less susceptible to image interactions and would generate a larger sample of adsorption and desorption events to draw conclusions from. Exploring higher Cl coverage would allow for the creation of more M-Cl systems which would provide further insight on the system's corrosion resistance. During the thermodynamic study we scanned the (T, P) domain but did not account for phase transitions of the bulk material. The most accurate results of the stability plots are reported around 800 $^\circ$C.
\section{Conclusion}
The surface corrosion resistance of Ni$_{70}$W$_{20}$Nb$_{10}$ was examined through the employment of density functional theory, a multi-cell Monte Carlo method, and thermodynamic modeling. The bulk structure was predicted to exhibit a single BCT phase closely resembling that of the NiW\textsubscript{4} structure at 800 \textsuperscript{o}C and 0 bar. The energetically favored (110) surface was used to conduct the surface study on. Incoming Cl showed a thermodynamic preference to Nb followed by W and Ni. From a PDOS analysis it was concluded that the alloy and chlorine s-, p-, and d-orbital states were hybridized and that strong electronic attraction did occur. The electronic insights agreed well with the reported high adsorption and desorption energies. The electronic structure of the surface slab was minimally perturbed by the adsorbed Cl atoms up to 5/15 ML, but it is anticipated that the PDOS curves would continue to change in response to higher concentrations of adsorbed Cl atoms.
\par 
The local corrosion resistance was examined for Ni, Nb, and W through the local adsorption of Cl to each species and the subsequent desorption of atomic and molecular Cl and the various metal-chloride molecules. Our findings suggested that Nb and W enhanced the corrosion resistance of Ni as their presence created regions that were thermodynamically preferred by incoming Cl and were less susceptible to Cl-facilitated dissolution from the surface model up to 5/15 ML. Ni, Nb, and W resisted Cl-induced dissolution from the surface model up to 5/15 ML indicating that all members of this alloy possessed superior resistance to localized surface degradation such as corrosive pitting. The strong resistance against Cl facilitated degradation exhibited by this alloy was in good agreement with experimental reports on Ni's corrosion resistance and the improved corrosion resistance of alloys containing either Nb, W, or both species in alloy samples. 
\par 
Using thermodynamic modeling we found that the stability of the chlorinated surface slab increased with the presence of Nb. This result was in good agreement with our findings from the adsorption events that Nb acted as a trapping sink for Cl on this surface slab. Overall, our results predicted that the (110) surface of Ni$_{70}$W$_{20}$Nb$_{10}$ possessed superior resistance to localized Cl-induced surface degradation. Additional surface studies need to be conducted before any generalizations can be made on the corrosion resistance of Ni$_{70}$W$_{20}$Nb$_{10}$. The findings of this work may guide the development of Ni-based alloys manufactured for molten salt containing systems, such as molten salt reactors. 
\section*{Data Availability}
Our implementation of the (MC)\textsuperscript{2} algorithm is available at \url{https://github.com/SaminGroup/Dolezal-MC2}. The data generated from the study is available at \url{https://github.com/SaminGroup/Dolezal-Nickel-Alloy}. The tool we developed for cutting surface slabs is available at \url{https://github.com/SaminGroup/Dolezal-SlabGenerator}.
\section*{Acknowledgements}
A.J. Samin would like to acknowledge funding from The Air Force Institute of Technology (AFIT) Faculty Research Council (FRC). In addition, the work was supported by computational allocations from the department of defense high performance computing (HPC) through Air Force Research Laboratory (AFRL) HPC Mustang and the Pittsburgh Supercomputing Center. 
\section*{Author Contributions}
\textbf{T.D. Dole\v{z}al} performed the calculations, developed code for process automation and data analysis, performed the data analysis, and prepared the manuscript. \textbf{A.J. Samin} advised and provided initial review the manuscript. Both authors contributed to the conceptualization and methodology of this work.

\bibliographystyle{ieeetr}

\end{document}